\documentstyle[12pt,psfig]{article}
\begin{document}
\begin{titlepage}

\centerline{{\bf QUASIPARTICLE EFFECTIVE MASS FOR THE TWO}}
\centerline{{\bf AND THREE DIMENSIONAL ELECTRON GAS \footnote{To appear
in {\it Phys.\ Rev.\ B\/} '96}}}
\centerline{Andrey Krakovsky \footnote{E-mail: krkvskya@acf2.nyu.edu}}
\centerline{\em Department of Physics, New York University, 
New York, New York 10003}
\centerline{Jerome K. Percus}
\centerline{\em Courant Institute of Mathematical Sciences and
Department of Physics,}
\centerline{\em New York University, New York, New York 10003}
\centerline{August 17 1995}
\begin{abstract}
We calculate the quasiparticle effective mass for the electron gas in
two and three dimensions in the metallic region.  We employ the single 
particle scattering
potential coming from the Sj\"{o}lander-Stott theory and enforce the
Friedel sum rule by adjusting the effective electron mass in a 
scattering calculation.  In 3D our effective mass 
is a monotonically
decreasing function of $r_s$ throughout the whole metallic domain, as 
implied by the most recent numerical
results.  In 2D we obtain reasonable agreement with the experimental
data, as well as with other calculations based on the Fermi liquid
theory.  We also present results of a variety of different treatments
for the effective mass in 2D and 3D.\\

\end{abstract}
\end{titlepage}
\section{Introduction}
The evaluation of the Fermi surface parameters has been a cornerstone of
the Fermi liquid theory since its early years. 
Precise knowledge of these parameters for an electron gas,
especially in the metallic domain, is not only a fundamental problem,
but is also extremely important for physical applications.  
At present, there is some
controversy about the three-dimensional results at metallic and
over-metallic densities. The two-dimensional
results are also of significant importance due to the recent nonvanishing
interest in the 2D physics stimulated by high-$T_c$
superconductivity, the fractional quantum Hall effect, as well as the
development of 2D electronic devices. 

As is well known from textbooks, quasiparticle excitations can be
characterized by the renormalization constant $Z(k_{\rm F})$ which is
related to the residue of the Green's function at the Fermi surface, and
the quasiparticle effective mass $m^{*}$.  In a simple-minded physical
picture, $1-Z(k_{\rm F})$ and $1-m^*/m$ both measure the amount of the
many-body effects in the electron gas.
In this paper we
will be concerned with the effective (renormalized) electron mass. We
will use the effective potential coming from the Sj\"{o}lander-Stott
theory \cite{ss}, and find the effective electron mass by 
adjusting the effective electron mass in a scattering calculation, so
that 
the Friedel
sum rule is satisfied.  Our approach is ``hydrodynamic" 
in  a sense that it does not
explicitly employ the microscopics of
the Fermi liquid, however, it requires the ``correct" static linear 
response function as
an empirical input .  Thus, the Fermi liquid character of the electron
gas will come in indirectly through the linear response which we use
in a parameterized form.  

Below we will outline the $GW$ format which is
a basis for the majority of calculations of the Fermi surface
parameters.  We will compare results of different approximation
schemes with ours.  For both 2D and 3D our results
are in a reasonable agreement with the most recent calculations based on
the Fermi liquid theory, as well as experimental data.  
\section{Effective mass in three dimensions}
In this section we will outline the $GW$ format for calculating the self
energy, and present numerical results for the 3D electron gas.
It is well known that in 3D the effective mass ratio $m^{*}/m$ is less than
unity in the high density limit.  The high density expansion ($r_s\ll
1$) was obtained in \cite{gell, dubois}.
In the metallic region ($1<r_s<8$) there has been some
controversy about the behavior of the effective mass ratio as a function
of the ground state density.  The formalism for evaluating the self
energy part was put forward by Hedin \cite{hedin} ($GW$ approximation).
In a more rigorous formulation \cite{ichimaru} it can be summarized as follows.
The standard starting point is the Dyson equation for the Green's
function:
\begin{equation}
	G_{\sigma}({\bf k}, \omega) = \frac{1}{\omega - \varepsilon_{\bf
k}^{(0)} - \Sigma_{\sigma}({\bf k}, \omega)}
	\label{eq:dyson}
\end{equation}
with $\varepsilon_{\bf k}^{(0)}$ the unperturbed energy and
$\Sigma_{\sigma}({\bf k}, \omega)$ the irreducible self energy.
The effective mass characterizes the quasiparticle excitation spectrum, 
\begin{equation}
	\varepsilon_{\bf k} = \frac{ {\bf k}^2}{2 m^{*}}\,,
\end{equation}
and in terms of the self energy is then given as:
\begin{equation}
        m^{*}/m=\left( 1- \frac{\partial \Sigma_{\sigma}({\bf k},
\omega)}{\partial \omega}\bigg|_{k=k_{\rm F}}\right)
        \left(1+\frac{m}{k}\frac{\partial \Sigma_{\sigma}({\bf
k},\omega)}{\partial k}\bigg|_{k=k_{\rm F}}\right)^{-1}\,,
        \label{eq:ratio}
\end{equation}
where $m$ is the unrenormalized (bare) electron mass.
The irreducible self energy can be approximately expressed in Dyson equation 
form as
\begin{equation}
	\Sigma_{\sigma}({\bf k}, \omega)=i 
	\int \frac{d{\bf q}}{(2\pi)^3}
	\int \frac{d\omega'}{2\pi} W({\bf q}, \omega')G_{\sigma}({\bf k}
	-{\bf q}, \omega - \omega')\,,
	\label{eq:108}
\end{equation}
where $W$ function incorporates the many-body effects. In general, it
can be approximated by
\begin{equation}
	W(q, \omega) = \frac{v_q}{\epsilon (q, \omega)}
	\Gamma (q, \omega)\,.
	\label{eq:w-function}
\end{equation} 
Here $v_q = 4 \pi e^2/ q^2$ is the (3D) bare Coulomb interaction,
$\epsilon (q, \omega)$ is the exact dielectric function, and $\Gamma
(q, \omega)$ is the vertex correction \cite{nakano.a}. 
In the work of Hedin \cite{hedin} the random phase approximation was
adopted by putting $\Gamma=1$ and using the RPA dielectric response in
(\ref{eq:w-function}):
\begin{equation}
	W^{\rm RPA} ( q, \omega)=v_{k}/  \epsilon^{\rm RPA} (
	q, \omega)\,.
	\label{eq:WRPA}
\end{equation}
Thus, the $W$-function was just an effective RPA screened interaction.
The self energy was obtained by
substituting the $W$-function (\ref{eq:WRPA}) into (\ref{eq:108})
and using the noninteracting Green's function in the right-hand-side of
(\ref{eq:108}).
Results of Hedin are shown in the Fig.~1 (black triangles).  The
effective mass ratio assumes its minimum at $r_s\approx 1$,
then increases in the metallic region,
and becomes greater than 1 for $r_s>3$.  More recent results indicate a
totally different behavior.  In the self-consistent approach of
Rietschel and Sham  \cite{rietschel} the
effective interaction function had the same form as in (\ref{eq:WRPA}),
but Eqs.\ (\ref{eq:dyson}) and (\ref{eq:108}) were solved self-consistently.
Their results (Fig.~1, empty triangles) indicate that the effective mass
ratio is a monotonically decreasing function of $r_s$ in the whole
metallic domain.  Similar results were obtained by Yasuhara and Ousaka
\cite{yasuhara} (Fig.~1, empty boxes) 
who analyzed the Landau interaction function \cite{nozieres}
using analytic fits based
on the Monte Carlo results of Ceperly and Alder \cite{ceperly}.  In
their work the decrease in the quasiparticle mass was due to both
spin-parallel and spin-antiparallel parts on the Landau interaction
function.  Another analysis of Landau interaction function based on SSTL
\cite{sstl} was carried out in \cite{pizzimenti}.  
Finally, a self-consistent scheme of Nakano and Ichimaru
with $W$-function incorporating vertex correction (\ref{eq:w-function})
produced a similar decreasing
behavior of the effective mass ratio (Fig.~1, circles).  
Their procedure is rather involved, we refer to \cite{nakano.b} for the details.

Our vehicle will now be the Sj\"{o}lander-Stott (SS) theory of
the two-component plasma \cite{ss}. As it was recently shown \cite{kra.a},
it is essentially a fluid description of the electron gas, and does not
directly employ
the microscopic structure of the Fermi liquid theory. It uses
the correct linear response of the electron gas as an empirical input
information. It is well known that this theory is capable of producing
reliable density profiles around a repulsive impurity. The density
profile
equation resulting from the SS theory is
\begin{equation}
        n_{q}^{{\rm ind, 3D}}=f^{\rm 3D}(q)\,\left(\,1+
        \frac{3}{4} \int_0^{\infty}
        k^2\,dk\,\left(\,1+{q^2-k^2\over 2qk}\,\ln{\left| \,{k+q\over
        k-q}\,\right|}\,\right)\,n_k^{{\rm ind, 3D}}\,\right)\,.
        \label{eq:profile3D}
\end{equation}
This gives the induced electron density $n_{q}^{{\rm ind}}$ around an
impurity of charge $Z$. Here, $f^{\rm 3D}(q)$ is the induced electron density in
the linear response approximation: 
\[
	f^{\rm 3D}(q) = \chi^{\rm 3D}(q) \frac{4 \pi Z e^2}{k_F^2 q^2}
\]
The static linear response $\chi^{\rm 3D}(q)$ is an input information and in our
calculations was
taken from the parameterization in \cite{farid}.
$k,q$ are in units
of the Fermi radius $k_F=(3\pi^2n)^{{1\over 3}}$, 
$n$ being the homogeneous ground state density. This profile relation
has the correct high--$q$ dependence which is responsible for satisfying
the cusp condition at the Coulomb source:
\begin{equation}
        n_q^{{\rm ind, 3D}}\bigg|_{q\rightarrow\infty}
        =Z\frac{16 \lambda r_s}{3\pi q^3}\,\left(
        1+\frac{n^{{\rm ind, 3D}}({\bf r}=0)}{n}\,\right)\,.
        \label{eq:cusp3D}
\end{equation}
The major shortcoming of the profile relation (\ref{eq:profile3D}) in the
domain of
repulsive impurities is overscreening by a hole. If the (repulsive)
impurity
charge is big enough the total electron density at the location of the
impurity goes negative. For most of the possible physical applications this
feature has an insignificant effect \cite{kra.b} because the region of
the non-physical behavior is very small and one can simply put the total
electron density to 0 in this region. However, to be on the
safe side, we restrict ourselves to the range of small enough
impurity charges 
\begin{equation}
	-0.3 < Z^{\rm 3D} < -0.03
	\label{eq:range3d}
\end{equation}
where the overscreening by a hole does not
occur and one
obtains a reliable density profile. We extract the effective single
particle scattering potential using the SS theory \cite{ss, kra.b}. 
With this scattering
potential we check the Friedel sum rule.

Friedel sum rule is a condition on the difference of the trace of the
logarithm of the single particle scattering matrix between the top and
the bottom of the band \cite{friedel, langer}. In our case of the
jellium model of electron gas it takes a familiar form
\begin{equation}
Z=\frac{2}{\pi}\sum_{l=0}^{\infty} n_l \delta_l(k_{\rm F})\,.
\label{eq:FSR}
\end{equation}
where the factor $n_l$ accounts for the angular degeneracy:
\[
	n_l = \left\{ \begin{array}{ll}
			2l+1 & \mbox{for 3D} \\
			2 - \delta_{l,0} & \mbox{for 2D}
			\end{array}
		\right.
\]
The factor of 2 comes from the spin degeneracy, $\delta_l(k_{\rm F})$
are
the partial wave scattering phase shifts at the Fermi momentum. This sum
rule has been routinely used to adjust free parameters of the
effective potential in a self-consistent fashion. In our case we take
the effective potential from the SS theory \cite{ss, kra.b}.  With this
effective potential we run a scattering calculation at the Fermi
surface, and, having obtained the phase shifts $\delta_l (k_{\rm F})$,
check the sum rule (\ref{eq:FSR}).  The particles that are scattered
at the Fermi surface are quasiparticles, not bare electrons.  Their mass
comes explicitly into the scattering calculation.  We adjust this
effective electron mass in the scattering calculation until (\ref{eq:FSR}) is
satisfied within 0.01\% accuracy.  This provides us with the value for
the effective mass. 
The procedure above is repeated for several values of $Z$ from the
region (\ref{eq:range3d}) in order to insure that the results are
independent of the impurity charge within a reasonable range

Our data  for 3D are plotted on Fig.~1 (black circles).  They clearly
indicate that the effective mass ratio is a monotonously 
decreasing function of the
Seitz radius throughout the whole metallic domain.  In the high density
limit they (as well as all the other data) converge to the 
limiting behavior \cite{gell, dubois}.  The agreement of our results with
the most recent 3D calculations suggests that the same procedure can be
tried in 2D since, in principle, the hydrodynamic model does not
distinguish between dimensions as long as the correct linear response is
employed.  Now, we will consider the situation in two dimensions.
\section{Effective mass in two dimensions}
Two dimensional calculations based on the many-body formalism 
have been carried out within the same
format of $GW$ approximation.  In the work of Jang and Min \cite{jang}
the $W$-function (\ref{eq:w-function}) is defined so that
the vertex correction $\Gamma =1$.  The dielectric function is expressed
in a standard way in terms of the local field correction ${\cal G}(q,
\omega)$:
\begin{equation}
	\epsilon(q, \omega)=1-\frac{v_q \chi_0(q, \omega)}{1+
	v_q {\cal G}(q, \omega) \chi_0 (q, \omega)}\,.
\end{equation}
The notation here is as before, but refers to the 2D quantities: $v_q =
2\pi^2 / q$, and the noninteracting response $\chi_0 (q, \omega)$ 
is as in \cite{stern}.  A
further approximation is conventionally made which consists of replacing
the dynamical local field correction by a frequency independent one: 
${\cal G}(q,
\omega) = {\cal G}(q)$.  Jang and Min employed different
parameterizations for the ${\cal G}(q)$ together with the
noninteracting Green's function in (\ref{eq:108}).  Their approach was
not self-consistent.
The employed approximations were: the RPA with ${\cal G}=0$;
the Hubbard approximation (HA) adopted to the 2D by Jonson
\cite{jonson} with
\begin{equation}
	{\cal G}^{\rm HA}(q)=\frac{1}{2}\frac{q}{\sqrt{q^2+k^2_{\rm F}}}\,,
\end{equation}
where $k_{\rm F} = \sqrt{2 \pi n}$ is the 2D Fermi radius; and the modified
Hubbard approximation (MHA) \cite{rice} with 
\begin{equation}
	{\cal G}^{\rm MHA}(q)= \frac{1}{2} \sqrt{q^2+k^2_{\rm F} + k^2_{\rm TF}}\,,
\end{equation}
where the Thomas-Fermi momentum is given by $k_{\rm TF}=2\pi n e^2/
\varepsilon_{\rm F} = \sqrt{2} r_s k_{\rm F}$.  The 2D Seitz radius is
given by $\pi (r_s a_0)^2 = 1/n$.
The effective mass ratios produced by these approximations are plotted
in Fig.~2: Hubbard approximation (empty triangles), modified Hubbard
approximation (empty boxes), and RPA (empty circles).

We proceed with the scheme we developed for the 3D case, but
adapted to 2D.  The profile
relation coming from the Sj\"{o}lander-Stott theory takes the form
\begin{equation}
        n_{q}^{{\rm ind, 2D}}=f^{{\rm 2D}}(q)\,\left(\,1+\int_0^
        {\infty}k\,dk\,{\Phi (\,q,k\,)\over
        \sqrt{k^2+q^2}}\,n_k^{{\rm ind, 2D}}\,\right)\,, 
	\label{eq:profile2D}
\end{equation}
where the notation is the same as in (\ref{eq:profile3D}), but refers to
2D.  $\Phi (q,k)$ results from
the angular integration \cite{ma}:
\begin{eqnarray}
        \Phi (q,k)&=&{q\over \pi}\left(\,{K\left({-2 a_{q,k}\over 1-a_{q,k}}
       \right) \over \sqrt{1-a_{q,k}}}+
        {K\left({2 a_{q,k}\over 1+a_{q,k}}
        \right) \over \sqrt{1+a_{q,k}}}\,\right)-{a_{q,k}k\over 4}\,
        _2F_1\left(\,{3\over 4},{5\over 4},2,a_{q,k}^2\right) \\
\mbox{with}\quad a_{q,k}&=&{2qk\over k^2+q^2} \quad . \nonumber
\end{eqnarray}
Here, $K(x)$ and $_2 F _1 (x)$ are the complete elliptic integral of the
first kind and the hypergeometric function, respectively.
The induced density in the linear response approximation $f^{{\rm
2D}}(q)$ is taken from the parameterization \cite{kra.b} of the numerical
results of Neilson et al.\  \cite{neilson}.  Just like
(\ref{eq:profile3D}), (\ref{eq:profile2D}) has the correct high--$q$
dependence which is responsible for satisfying the cusp condition at the
Coulomb source:
\begin{equation}
        n_q^{{\rm ind, 2D}}\bigg|_{q\rightarrow\infty}
        ={Z\,2\sqrt{2}\,r_s\over q^3}\,\left(
        1+{n^{{\rm ind, 2D}}({\bf r}=0)\over n}\,\right)\quad .
        \label{eq:cusp2D}
\end{equation}
In order to avoid dealing with the overscreening by a hole we take the
repulsive impurity charge to be small enough 
\begin{equation}
	-0.1 < Z^{\rm 2D} < -0.01 
	\label{eq:range2d}
\end{equation}
As in the
previous section, we extract the effective potential using the SS theory and
run a 2D scattering calculation at the Fermi momentum adjusting the
effective electron mass until the Friedel sum rule (\ref{eq:FSR})
is satisfied.  Just as in 3D, we repeat the calculation for different
values of $Z$ within the range (\ref{eq:range2d}).
Our results for the effective mass ratio are shown on Fig.~2.
They fit roughly in between the results of the HA and the MHA.  The HA
is known to take into account the exchange interaction and neglect
correlations, while the MHA is an attempt to incorporate correlations.
To compare with the experimental data we plot the magnetoconductivity
measurements of Smith and Stiles \cite{smith} (Fig.~2, crosses).  The agreement
of our results with the experimental values is rather reasonable.  We
also plot the high density expansion (Fig.~2, dashed line) obtained by 
Isihara and Toyoda \cite{isihara} from the specific heat calculation:
\begin{equation}
	m^{*}/ m = 1+0.043 r_s\,.
	\label{eq:limit2D} 
\end{equation}
This result was obtained by including first- and second-order exchange, as well as the ring
diagram contributions for small but finite temperatures.

And last, a few words about our numerical procedure would be appropriate.
In the
solution of (\ref{eq:profile3D}) and (\ref{eq:profile2D}) we could use
(\ref{eq:cusp3D}) and (\ref{eq:cusp2D}), respectively, to
approximate the high--$q$ behavior, and after solving in the remaining
domain match the solutions. This procedure is, however, very tedious
and ineffective. We found that the solution is not affected if we solve
the integral equation on the whole semiinfinite domain $[0, \infty)$
using the Gauss-rational rule for discretizing the integral.  Of course,
2D calculations require care because of weaker convergence of
integrals.  In the
scattering calculation we used 12 partial waves.
\section{Conclusion}
We have presented our results for the electron effective mass ratio using the
effective potential coming from the SS theory for two and three
dimensions.  The
calculations were carried out for small repulsive impurities where
the SS works.  The effective mass was extracted by enforcing the
Friedel sum rule in the scattering calculation for the effective
potential.  We compared our results with
the most recent numerical, as well as experimental data.  

As for the 3D results (Fig.~1), there has been some uncertainty in the
behavior   
of $m^{*} / m$ in the metallic domain.  Earlier results predicted that
this ratio should increase with $r_s$, while more recent results
indicate
that it is a decreasing function of $r_s$.  In our treatment the
effective mass ratio is a monotonically decreasing function of the Seitz
radius in the whole metallic domain.

In 2D (Fig.~2) we present our results together with the results based on the
$GW$-approximation.  Our treatment seems to fall ``in between" the
Hubbard and the modified Hubbard approximations.  Our
results give reasonable agreement with the experimental data.

In conclusion, it is
interesting to relate the 2D and 3D results between themselves.  It is
well known that the exchange interaction diminishes the effective mass,
while correlations shift it in the opposite direction.  The quasiparticle
effective mass is a result of an interplay between the exchange and correlation
contributions.  It comes as no surprise that because correlations are 
stronger in 2D than in 3D, generally, the effective mass is greater in 2D
than in 3D for the same value of $r_s$.  
\section*{Acknowledgments}
The authors would like to thank Professor I. Nagy for providing some of the
most important
references for this work; Professor J. L. Birman for extremely helpful
discussions.  Partial support of the National Science
Foundation and the Physics Department at New York University is
gratefully acknowledged.

\vfill\eject
\section*{Figure Captions}
\noindent Fig.\ 1.  The effective mass ratio $m^*/m$ vs.\ the Seitz
radius $r_s$ for the 3D electron gas.  Results of Hedin \cite{hedin} (black
triangles), Nakano and Ichimaru \cite{nakano.b} (empty circles), Yasuhara
and Ousaka \cite{yasuhara} (empty boxes), present (dark circles), and
Rietschel and Sham \cite{rietschel} (empty triangles).  The dashed line is
the high density limit \cite{gell, dubois}.  Our results
indicate that the effective mass is a monotonically decreasing function
of $r_s$ throughout the whole metallic domain.\\

\noindent Fig.\ 2.  The same as in Fig.\ 1.  Results of Jang and Min
\cite{jang} in the Hubbard approximation (empty triangles), modified
Hubbard approximation (empty boxes), RPA (empty circles); present theory
(black circles); experimental data of Smith and Stiles \cite{smith} 
(crosses).  The
dashed line is the high density result (\ref{eq:limit2D}) of Isihara and
Toyoda \cite{isihara}.
\vfill\eject
\begin{figure}[htbp]
\psfig{file=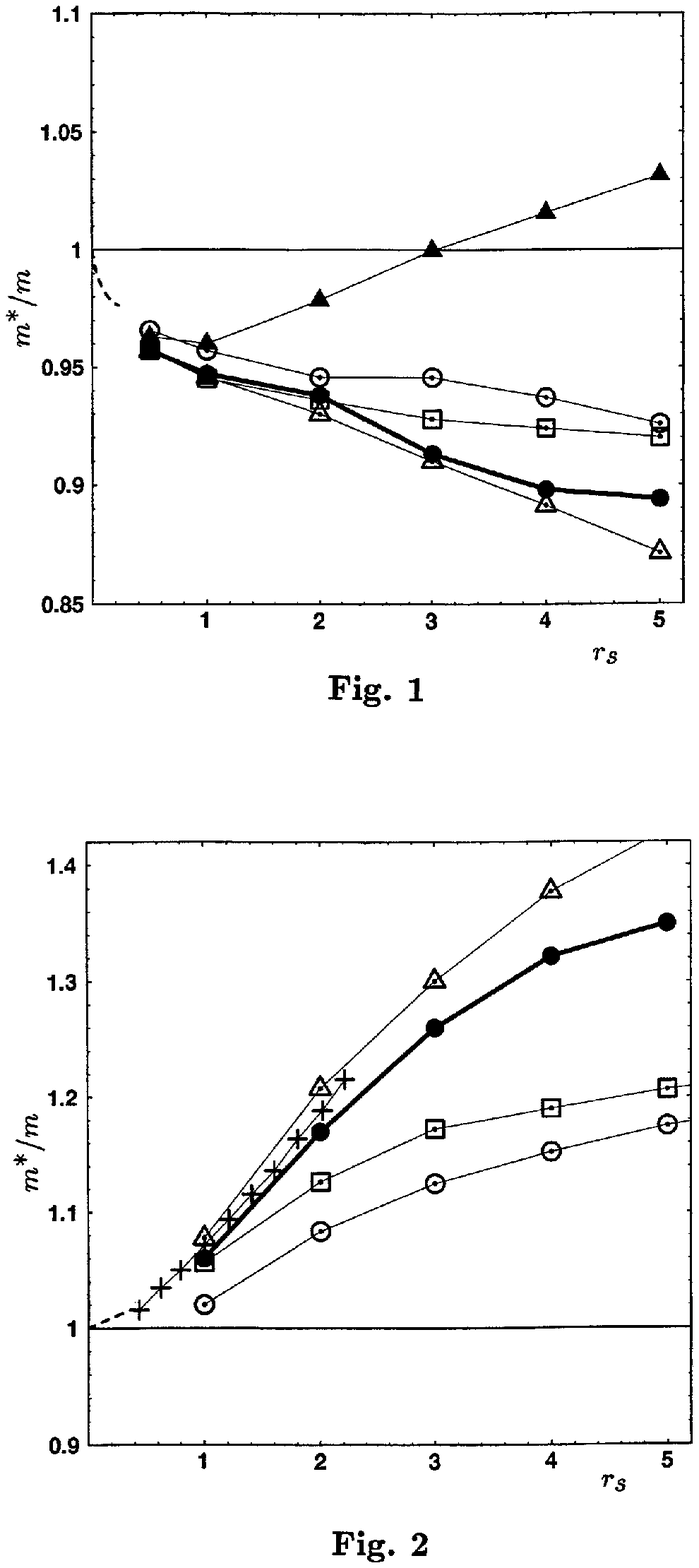}
\end{figure}

\begin{thebibliography}{99}
\bibitem{ma} We used Mathematica 2.1 symbolic integration packages.
\end{thebibliography}
\end{document}